# Quantum Convolution for Structure-Based Virtual Screening

Pei-Kun Yang
E-mail: peikun@isu.edu.tw
ORCID: https://orcid.org/0000-0003-1840-6204

**Abstract**

Structure-based virtual screening (SBVS) is a key computational strategy for identifying potential drug candidates by estimating the binding free energies ($\Delta G_{bind}$) of protein–ligand complexes. The immense size of chemical libraries, often exceeding billions of compounds, combined with the need to account for protein and ligand conformations and ligand translations and rotations, renders such tasks computationally intensive on classical hardware. This study proposes a quantum convolutional neural network (QCNN) framework to estimate $\Delta G_{bind}$ efficiently. Using the PDBbind v2020 dataset, with the core set designated as the test set and the remaining complexes used for training, we trained QCNN models with 9 and 12 qubits. The best-performing model achieved a Pearson correlation coefficient of 0.694 on the test set. To assess robustness, we introduced quantum noise under two configurations. While noise increased root mean square deviation, the Pearson correlation coefficient remained largely stable. These results demonstrate the feasibility and noise tolerance of QCNNs for high-throughput virtual screening and highlight the potential of quantum computing to accelerate drug discovery.

**Introduction**

Identifying small molecules that inhibit protein activity is critical for developing therapies targeting various diseases. Structure-Based Virtual Screening (SBVS) has emerged as a powerful computational strategy for discovering potential drug candidates by leveraging the three-dimensional structures of proteins and small molecules. By estimating binding free energies ($\Delta G_{bind}$) or binding scores, SBVS enables the prioritization of high-affinity ligands for subsequent experimental validation. This approach significantly accelerates drug discovery while reducing costs, especially given the vast chemical space. Even when limited to small, drug-like molecules, the estimated chemical space exceeds $10^{60}$ compounds [1, 2]. Several publicly available ligand libraries have been curated to facilitate virtual screening [3-5]. For instance, the ZINC22 database provides access to approximately 4.5 billion purchasable compounds, offering a valuable resource for drug discovery research [6].

Accurate estimation of $\Delta G_{bind}$ has traditionally relied on statistical thermodynamics and quantum chemistry, which, while theoretically rigorous, are computationally demanding [7-9]. In practical applications, molecular dynamics simulations are employed to sample the conformational space of molecular systems. These simulations often incorporate explicit solvent models to more realistically capture intermolecular interactions and have demonstrated good agreement with experimental measurements [10]. However, applying such methods to large-scale ligand libraries remains impractical due to the required computational resources.

Researchers have developed simplified models based on electrostatic and van der Waals interactions to address these limitations, approximating entropic and solvent effects. These approaches often substitute $\Delta G_{bind}$ with binding scores, thereby enabling more computationally efficient screening processes [11-15]. Nevertheless, screening billions of compounds still requires molecular docking algorithms to generate three-dimensional protein-ligand complexes. Although docking methods are generally effective, they frequently introduce structural deviations of approximately 2 Å, which can significantly affect binding energy estimations due to the sensitivity of electrostatic and van der Waals interactions to atomic coordinates [16-19].

Recent machine learning (ML) advances have further enhanced SBVS. For example, models such as Gnina represent protein-ligand complexes as four-dimensional occupancy grids that encode both atomic types and spatial information. This representation helps mitigate errors introduced by docking and reduces the sensitivity of scoring functions to small perturbations in atomic coordinates. Additionally, ML-based methods provide greater flexibility in parameter tuning, allowing them to capture complex molecular interactions, including quantum mechanical effects that are challenging to model explicitly with conventional approaches [20-26].

Quantum computing introduces a fundamentally new paradigm for SBVS by exploiting quantum superposition and entanglement to achieve enhanced computational efficiency. Unlike classical systems, where $n$ bits represent $n$ discrete states, quantum systems with $n$ qubits can simultaneously encode $2^n$ quantum states. This exponential scaling enables quantum computers to process large datasets with unprecedented parallelism. For instance, a single quantum logic gate can simultaneously act on all basis states. With $n$ qubits, the binding free energy of a single protein-ligand complex can be computed; by adding $m$ qubits, the binding free energies of $2^m$ ligands can be calculated in parallel; and with additional $p$ qubits, both position and orientation for $2^p$ ligand configurations can be evaluated simultaneously. Such scalability far exceeds the capabilities of classical systems [27-32].

Recent advances in quantum machine learning (QML) have sparked growing interest in its potential applications for drug discovery. Early studies have proposed hybrid quantum-classical models, including variational quantum circuits and quantum support vector classifiers, for molecular classification tasks [33, 34]. In addition, quantum generative approaches such as hybrid quantum generative adversarial networks have been explored for designing drug-like molecular structures [35]. While these efforts demonstrate the feasibility of applying QML in cheminformatics, their integration into SBVS remains in its early stages.

This study proposes a quantum convolutional neural network (QCNN) approach for SBVS to estimate $\Delta G_{bind}$ values of protein-ligand complexes. Due to the limited accessibility and high cost of current quantum hardware, all computations are conducted using CPU/GPU-based quantum circuit simulators. It is important to emphasize that conventional machine learning algorithms are highly optimized for classical hardware, achieving significantly greater computational efficiency. In contrast, simulating quantum algorithms on classical systems is inherently less efficient than executing algorithms natively designed for CPUs or GPUs. Therefore, the results presented in this study should be interpreted as a proof of concept for an algorithm intended for future quantum computers, rather than a direct performance comparison with classical methods. Furthermore, QCNNs face intrinsic limitations, such as the absence of nonlinear activation functions and the frequent emergence of barren plateau issues during parameter optimization. Accordingly, this work aims to establish a preliminary framework for QCNN-based SBVS, serving as a foundation for future refinement and advancement.

**Methods**

**Converting Protein Structures into Occupancies.** This study employed the PDBbind v2020 dataset, which contains 19,443 protein-ligand complexes with experimentally determined $pK_d$ values. The Core set, consisting of 285 complexes, was designated as the test set, while the remaining 19,158 complexes were used for training [15, 25, 33].

The geometric center of each ligand was aligned with the center of a cubic grid measuring 16 Å per side to ensure consistent spatial orientation. Atoms were classified into eight categories based on chemical properties: carbon (C), nitrogen (N), oxygen (O), and others, with separate distinctions for protein and ligand atoms. Assigned van der Waals radii were 1.9 Å for C, 1.8 Å for N, 1.7 Å for O, and 2.0 Å for all other non-hydrogen atoms. Spatial occupancies were calculated using Equation (1), which is based on the ratio $r$ between the distance from the atomic center and its van der Waals radius [36].

$$O_{atom}(r) = \begin{cases} e^{-2r^2} & r < 1 \\ \left(\dfrac{3-2r}{e}\right)^2 & 1 \leq r < 1.5 \\ 0 & r \geq 1.5 \end{cases} \quad (1)$$

Each protein-ligand complex was initially represented as a 32 × 32 × 32 voxel grid, which was subsequently downsampled to either 8 × 8 × 8 or 4 × 4 × 4 using max-pooling. This process yielded 4,096 or 512 occupancy values per complex, computed by summing the contributions ($O_{atom}$) of all atoms belonging to the same type. Occupancies were normalized to ensure compatibility with quantum computational frameworks, such that the sum of their squared values equaled 0.5, separately for protein and ligand atoms.

Experimentally measured p$K_d$ values were converted to Δ$G_{bind}$ values (kcal/mol) using Equation (2), thereby providing a standardized dataset for training and evaluating the QCNN model.

$$\Delta G_{bind} = -\ln 10 * RT * pK_d \tag{2}$$

**Quantum Circuit Design.** We developed a quantum circuit to estimate Δ$G_{bind}$ values for protein-ligand complexes, encoding each complex into a quantum state using either 9 or 12 qubits (Figure 1). A series of quantum filters (Qfilters) were applied to extract hierarchical features from the encoded quantum states. These quantum features were subsequently transformed into classical data via projective measurement. Specifically, qubit 0 was measured to determine the probabilities of observing the |0⟩ and |1⟩ states. These measurement probabilities were then weighted by trained parameters to predict the Δ$G_{bind}$ value.

Qfilters were implemented as parameterized unitary matrices. Singular Value Decomposition (SVD) was applied to each matrix to satisfy the unitary constraint imposed by quantum mechanics. Specifically, each parameterized matrix was decomposed into U, Σ, and V* components, and a valid unitary matrix was reconstructed by combining U and V*, omitting the singular values in Σ [37]. This method ensures that Qfilters strictly comply with the physical principles of quantum systems, thereby enabling accurate and efficient modeling of Δ$G_{bind}$.

Although Qfilters can alternatively be constructed using parameterized quantum gates and controlled operations, such designs typically result in deeper circuits. Increased circuit depth significantly raises the computational cost during classical simulation and exacerbates the barren plateau problem, wherein gradients vanish and impede effective training [38]. In contrast, implementing Qfilters as parameterized unitary matrices substantially reduces circuit depth, lowers computational overhead, and helps mitigate the barren plateau issue.

Since quantum measurements are inherently probabilistic, multiple shots are required to obtain reliable estimates of measurement probabilities. Reducing the number of qubits measured decreases the required shots, lowering computational overhead. Accordingly, this study minimized the number of measured qubits to one, with only a single qubit measured per protein-ligand complex.

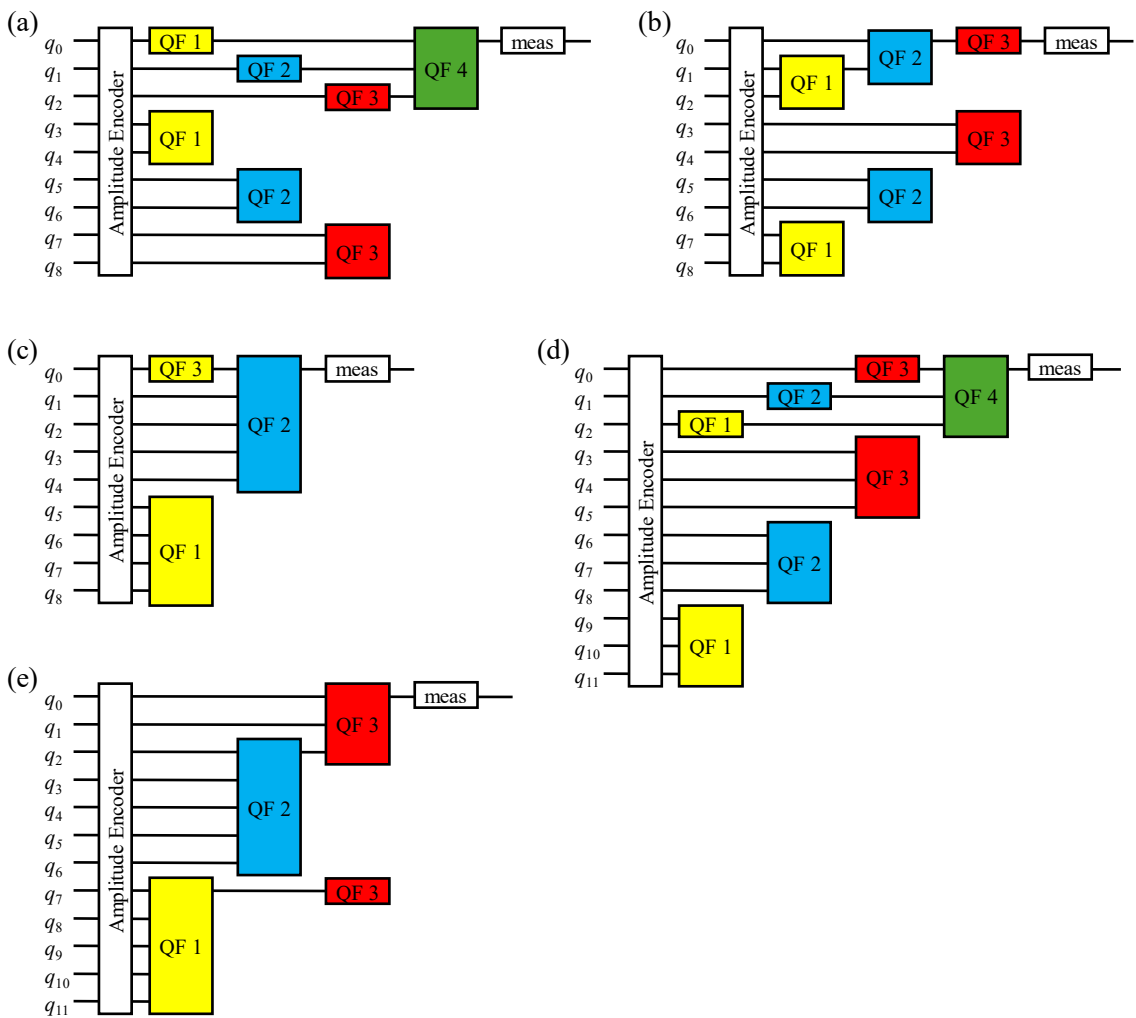

**Figure 1.** QCNN architectures with 9 and 12 qubits. Each protein–ligand complex is encoded into a quantum state using amplitude encoding and processed through a hierarchical series of quantum filters for feature extraction. The classical output is obtained by measuring qubit 0. Subfigures (a)–(c) show 9-qubit architectures with filters acting on up to 3, 4, and 5 qubits, respectively, while (d) and (e) depict 12-qubit architectures using filters operating on up to 4 and 5 qubits, respectively, to capture deeper quantum correlations.

**Model Implementation and Training Details.** The QCNN for $\Delta G_{bind}$ prediction was implemented using PyTorch. Model parameters consisted of Qfilters designed to extract hierarchical features from the input occupancy grids. Training was carried out using stochastic gradient descent optimization. The input occupancy data were sequentially transformed through a series of learned unitary operations. After each transformation, the resulting amplitudes were reshaped to preserve spatial structure. The amplitude values were squared and summed after the final transformation to produce classical, probability-like features. These features were then linearly combined with trainable weights to predict $\Delta G_{bind}$ values. The mean squared error between predicted and reference values served as the loss function for model optimization.

The training was conducted using learning rates of $10^{-2}$, $10^{-3}$, $10^{-4}$, or $10^{-5}$, with a momentum coefficient 0.9. The batch size was fixed at 32 for both the training and evaluation phases. Depending on the experimental configuration, each protein-ligand complex was represented using either 9 or 12 qubits. The model architecture consisted of multiple sequential Qfilters, each implemented as a parameterized unitary matrix with initial values randomly drawn from a uniform distribution over the range [0, 1].

The training was performed for 10,000 iterations. In each iteration, input occupancy grids were processed in batches, and model parameters were updated to minimize the mean squared error loss. After training, model performance was assessed on an independent test set using the same preprocessing pipeline and batch-processing protocol.

**Quantum Circuit Validation and $\Delta G_{bind}$ Prediction.** This module was developed to validate whether the model parameters trained in PyTorch are compatible with quantum hardware constraints. The trained parameters were loaded from saved checkpoints and used to construct a quantum circuit in PennyLane. In the quantum circuit, input occupancy vectors were initialized using the QubitStateVector function, and hierarchical transformations were applied via sequential Qfilters. A projective measurement was performed on qubit 0 to obtain the probabilities of observing the $|0\rangle$ and $|1\rangle$ states. These probabilities were then combined using trained linear weights to predict each complex's $\Delta G_{bind}$ value. Predicted $\Delta G_{bind}$ values obtained from the quantum circuit simulation were compared with those generated by the original PyTorch implementation to verify the physical validity and consistency of the learned parameters. This validation step ensures that the trained model can, in principle, be executed on real quantum hardware without violating the unitary constraints imposed by quantum mechanics.

**Noisy Quantum Circuit Evaluation.** To assess the robustness of the trained QCNN models under realistic quantum noise, we conducted noisy quantum circuit simulations using PennyLane's default.mixed simulator. This simulator supports mixed-state evolution and enables evaluation under realistic decoherence conditions. All five QCNN architectures presented in Fig. 1 were evaluated using the pre-trained parameters obtained from noise-free training.

Each protein-ligand complex was encoded into a 9- or 12-qubit quantum state via amplitude encoding using the QubitStateVector function. The quantum state was then processed through a sequence of Qfilters for hierarchical feature extraction, followed by measurement on qubit 0 to generate the predicted $\Delta G_{bind}$ value.

We implemented and compared two noise injection strategies: Final-Qubit Noise and Layer-Wise Noise. In the Final-Qubit Noise configuration, depolarizing noise (probability = 0.05) and phase damping noise (probability = 0.03) were applied exclusively to qubit 0 after the final Qfilter block and immediately before measurement. This setting simulates measurement-related decoherence effects on the output qubit. In contrast, the Layer-Wise Noise configuration introduced noise throughout the circuit. The same depolarizing and phase damping channels were applied after each Qfilter block to all qubits involved in the respective operation, thereby modeling the cumulative impact of gate noise and decoherence across the entire circuit.

By evaluating predictive performance under both noise models, we aimed to examine how the location and accumulation of quantum noise affect the stability and accuracy of the QCNN architectures.

**Results**

**Performance Comparison of QCNN Architectures.** Table I summarizes the predictive performance of five QCNN architectures (Fig. 1), which differ in qubit count, Qfilter configuration, and the number of trainable parameters. The models were trained on the PDBbind v2020 dataset and evaluated using root mean square deviation (RMSD) and Pearson correlation coefficient (PCC) on both training and test sets.

Among the 9-qubit models, architecture 1c (Qfilters: 5+5) achieved the best test performance, with an RMSD of 2.23 kcal/mol and a PCC of 0.693. It outperformed 1a (3+3+3+3) and 1b (4+4+3), which showed lower correlations and slightly higher deviations. Notably, 1c also possessed the largest number of parameters among the 9-qubit models.

For the 12-qubit models, architecture 1f (Qfilters: 4+4+4+3) yielded the highest test PCC of 0.694 and a relatively low test RMSD of 2.27 kcal/mol, demonstrating strong generalization. In contrast, although architecture 1g (Qfilters: 5+5+4) had more parameters (2,306 total, 1,114 independent) and achieved better training results (RMSD: 2.03 kcal/mol, PCC: 0.653), its test performance degraded (PCC: 0.677; RMSD: 2.36 kcal/mol), suggesting overfitting.

Across all models, the standard deviations of RMSD and PCC remained small (e.g., ≤0.06 for RMSD and ≤0.017 for PCC on the test set). As shown in Table I, this low variance indicates that learning rate and training stochasticity had minimal effects on model stability, underscoring the robustness and reproducibility of the QCNN approach under varying hyperparameters.

Figure 2 presents scatter plots comparing predicted versus experimental $\Delta G_{bind}$ values for five QCNN architectures (Fig. 1a–e), with each model evaluated under different learning rates. The plots display the runs with the highest and lowest PCC values for each architecture.

Figure 3 illustrates RMSD trajectories over 10,000 training epochs for the same five architectures, evaluated at four learning rates: $10^{-2}$ (blue), $10^{-3}$ (green), $10^{-4}$ (red), and $10^{-5}$ (orange). All models exhibit an initial fluctuation phase followed by stabilization. Learning rate $10^{-2}$ consistently resulted in larger oscillations, while $10^{-4}$ and $10^{-5}$ showed smoother convergence than $10^{-2}$. Despite the differences, RMSD remained within a narrow range, and curve shapes were consistent across architectures.

**Table I.** Summary of quantum circuit configurations and predictive performance corresponding to the architectures shown in Fig. 1. "N qubit" denotes the number of qubits used in each circuit. "Qfilter" specifies the size and sequence of quantum filters applied in each architecture. "N_par" indicates the total number of trainable parameters, with the number of independent parameters shown in parentheses. RMSD and PCC values are reported for both the training and test sets. Each value represents the mean ± standard deviation from four independent training runs, using learning rates of $10^{-2}$, $10^{-3}$, $10^{-4}$, and $10^{-5}$.

| QC_Fig | N_qubit | Qfilter | N_par | Train | | Test | |
|---|---|---|---|---|---|---|---|
| | | | | RMSD | PCC | RMSD | PCC |
| 1a | 9 | 3+3+3+3 | 258 (86) | 2.18 ±0.08 | 0.544 ±0.001 | 2.37 ±0.02 | 0.653 ±0.007 |
| 1b | | 4+4+3 | 578 (270) | 2.10 ±0.03 | 0.573 ±0.002 | 2.37 ±0.06 | 0.655 ±0.017 |
| 1c | | 5+5 | 2,050 (994) | 1.99 ±0.00 | 0.622 ±0.001 | 2.23 ±0.05 | 0.693 ±0.013 |
| 1f | 12 | 4+4+4+3 | 834 (390) | 2.05 ±0.02 | 0.594 ±0.003 | 2.27 ±0.05 | 0.694 ±0.012 |
| 1g | | 5+5+4 | 2,306 (1,114) | 2.03 ±0.17 | 0.653 ±0.004 | 2.36 ±0.22 | 0.677 ±0.010 |

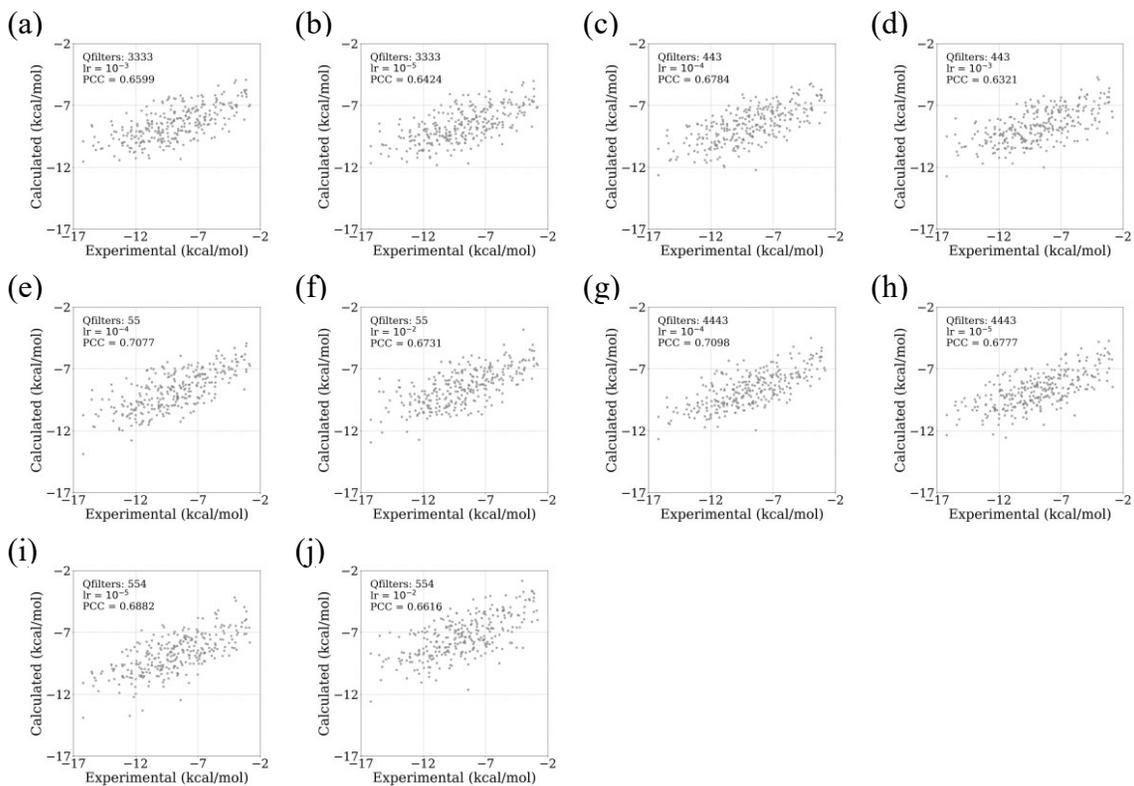

**Figure 2.** Predicted versus experimental ΔG$_{bind}$ values on the test set using QCNN architectures are shown in Fig. 1. Subplots (a)–(j) correspond to architectures from Fig. 1a–e, using Qfilter configurations of 3333, 443, 55, 4443, and 554, respectively. The results with the highest and lowest Pearson correlation coefficients (PCC) under varying learning rates are displayed for each configuration. "Qfilters" represents the filter configuration, "lr" denotes the learning rate, and "PCC" indicates the correlation between predicted and experimental binding free energies.

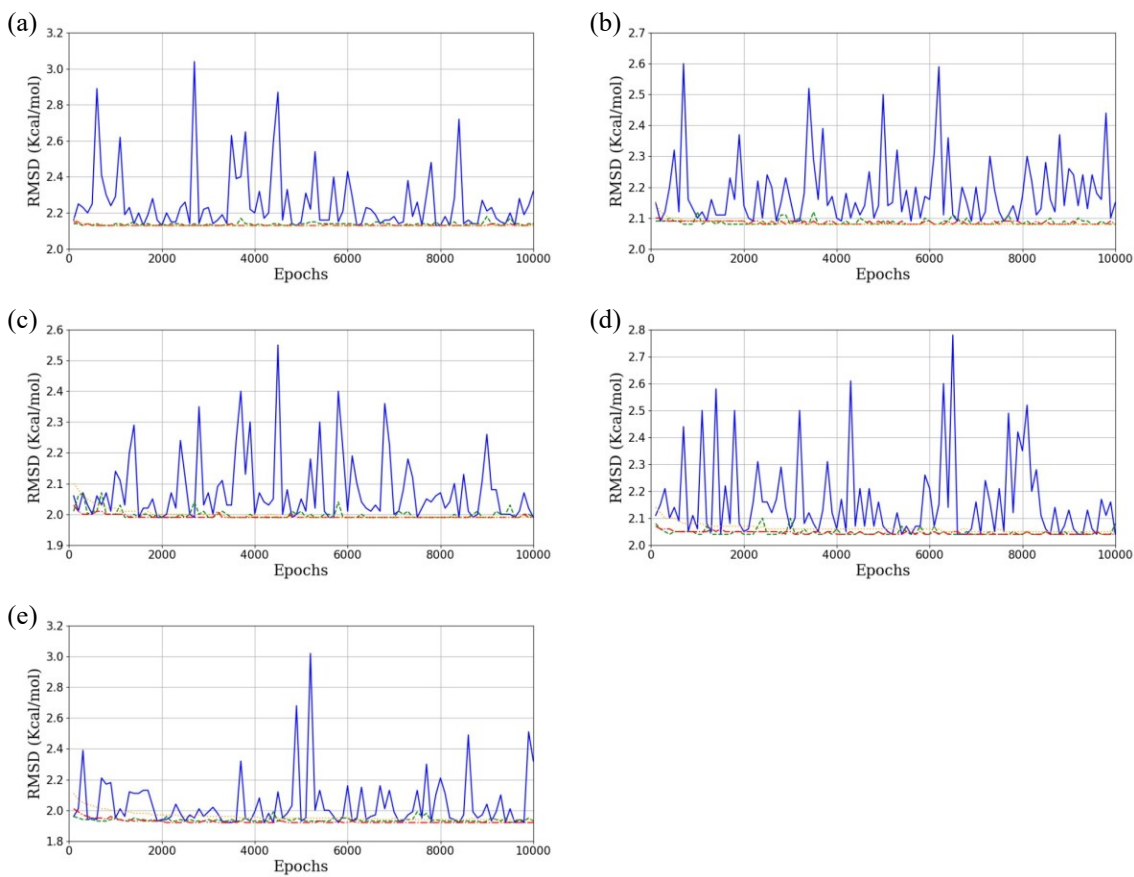

**Figure 3.** RMSD between predicted and experimental ΔG$_{bind}$ values during training over 10,000 epochs, using different learning rates: $10^{-2}$ (blue), $10^{-3}$ (green), $10^{-4}$ (red), and $10^{-5}$ (orange). Subplots (a)–(e) correspond to the QCNN architectures shown in Fig. 1a–e, respectively.

**Performance Under Noisy Quantum Conditions.** Table II summarizes the predictive performance of five QCNN architectures (Fig. 1a–e) evaluated under two quantum noise settings: Final-Qubit Noise and Layer-Wise Noise. RMSD and PCC values were computed for both training and test datasets in each setting. All results are reported as mean ± standard deviation across four learning rates.

Under the Final-Qubit Noise condition, test RMSD values ranged from 2.37 to 2.51 kcal/mol, and test PCC values spanned 0.653 to 0.694. Training RMSD values fell between 2.06 and 2.20 kcal/mol, while training PCC values ranged from 0.544 to 0.652. However, architecture 1.e exhibited a notably higher training RMSD deviation (±0.30), suggesting sensitivity to initialization or hyperparameter choices.

In the Layer-Wise Noise setting, test RMSD values ranged from 2.52 to 2.93 kcal/mol, and test PCC values from 0.652 to 0.694. Training RMSD ranged from 2.16 to 2.53 kcal/mol, and training PCC from 0.543 to 0.621, with generally moderate standard deviations (mostly ≤0.17 for RMSD and ≤0.016 for PCC), except for some outliers.

These results suggest that while most architectures maintained stable performance across different learning rates and random seeds, some configurations (e.g., 1.e) are more sensitive to training variability and may require further regularization or architectural refinement.

**Table II.** Predictive performance of five QCNN architectures under two quantum noise models: Final-Qubit Noise and Layer-Wise Noise. Architectures (1.a–1.e) correspond to those in Fig. 1. Each model was evaluated using pre-trained parameters obtained from noise-free training and tested under both noise settings. In the Final-Qubit Noise configuration, depolarizing noise (p = 0.05) and phase damping noise (p = 0.03) were applied to qubit 0 after the final Qfilter block. In the Layer-Wise Noise configuration, the same noise types were applied after each Qfilter layer to all qubits involved in the respective operation. RMSD and PCC values are reported as mean ± standard deviation, averaged over four learning rates and multiple random seeds.

| QC_Fig | Final-Qubit Noise | | | | Layer-Wise Noise | | | |
|---|---|---|---|---|---|---|---|---|
| | Train | | Test | | Train | | Test | |
| | RMSD | PCC | RMSD | PCC | RMSD | PCC | RMSD | PCC |
| 1.a | 2.20 ±0.02 | 0.544 ±0.001 | 2.46 ±0.05 | 0.653 ±0.007 | 2.44 ±0.17 | 0.543 ±0.001 | 2.81 ±0.21 | 0.652 ±0.007 |
| 1.b | 2.16 ±0.04 | 0.573 ±0.002 | 2.50 ±0.08 | 0.655 ±0.017 | 2.53 ±0.16 | 0.572 ±0.002 | 2.93 ±0.19 | 0.658 ±0.016 |
| 1.c | 2.06 ±0.04 | 0.622 ±0.001 | 2.38 ±0.03 | 0.693 ±0.013 | 2.16 ±0.06 | 0.621 ±0.001 | 2.52 ±0.04 | 0.694 ±0.014 |
| 1.d | 2.09 ±0.04 | 0.594 ±0.003 | 2.37 ±0.06 | 0.694 ±0.012 | | | | |
| 1.e | 2.14 ±0.30 | 0.652 ±0.004 | 2.51 ±0.34 | 0.677 ±0.010 | | | | |

**Discussions**

**The Potential of Quantum Computing in Structure-Based Virtual Screening.** In SBVS, the $\Delta G_{bind}$ of protein–ligand complexes is computed to identify small molecules with the lowest $\Delta G_{bind}$ as potential drug candidates. However, this screening process poses significant computational challenges. The chemical space of drug-like molecules is extraordinarily vast, with estimates exceeding $10^{60}$ compounds. Moreover, one must account for conformational flexibility in both proteins and ligands and the translational and rotational positioning of the ligand within the binding pocket. Accurate $\Delta G_{bind}$ estimation via statistical thermodynamics and quantum chemistry is computationally intensive, rendering exhaustive evaluations infeasible even with today's most powerful classical computing resources. Thus, efficient and predictive screening methods are essential for the practical application of SBVS.

To manage the complexity of the screening process, researchers have developed conventional approaches that simplify the problem by considering a limited number of protein and ligand conformations. Docking algorithms are used to identify plausible binding poses, and instead of first-principles $\Delta G_{bind}$ calculations, empirical or statistical scoring functions are applied to estimate binding affinity. Although these approximations do not guarantee strong experimental binding, they increase the probability of discovering high-affinity candidates compared to random selection.

Because $\Delta G_{bind}$ calculations across different protein–ligand pairs are inherently parallelizable, GPUs are commonly used to accelerate these tasks. Nevertheless, even massive GPU parallelism cannot explore the vast configuration space. Quantum computing offers a promising alternative by leveraging quantum superposition to achieve exponential parallelism. In principle, if $n$ qubits represent a single protein–ligand complex and $m$ additional qubits encode binding configurations, a quantum computer could simultaneously represent $2^m$ combinations. For example, with $m = 300$, approximately $10^{90}$ configurations could be evaluated in parallel, encompassing ligand diversity, conformations, and spatial arrangements.

Several approaches can be employed to compute $\Delta G_{bind}$ within a quantum framework. Classical force-field-based methods, including electrostatics and van der Waals interactions, are atomistic and well-suited for quantum parallelization. Solvation energy estimation, which involves solving the Poisson–Boltzmann equation, may also benefit from quantum algorithms. Additionally, quantum chemistry methods that model electronic interactions from first principles align naturally with quantum computation and could further enhance predictive accuracy.

Despite this potential, current quantum hardware cannot process datasets as large as $10^{60}$ ligands or store the associated quantum state space. Instead, using unitary matrices, quantum algorithms must be designed to generate ligand structures, conformations, and spatial transformations dynamically. These operations include simulating ligand translation and approximating rotation within quantum circuits. Realizing a practical quantum SBVS framework will require interdisciplinary collaboration across quantum computing, machine learning, bioinformatics, and computational chemistry.

**Estimating $\Delta G_{bind}$ of Multiple Protein–Ligand Complexes Simultaneously.** We developed a quantum computational framework to estimate $\Delta G_{bind}$ for multiple protein–ligand complexes in parallel (Figure 4). The circuit utilizes $n$ qubits to encode molecular interaction features of a single protein–ligand complex. In contrast, additional $m$ qubits enable simultaneous computation for up to $2^m$ complexes within a single quantum execution.

The QC$_{bind}$ module calculates the $\Delta G_{bind}$ between a protein and a single ligand. Incorporating $m$ Identity gates (I) distributes the QC$_{bind}$ operation across multiple complexes, simultaneously enabling the computation of binding free energies between a protein and $2^m$ ligands by leveraging quantum parallelism. The measurement step collapses quantum states into classical data, which are subsequently processed to generate final predictions.

Equation (3) governs the relationship between the input quantum states (I) and the output quantum states (O) of the $n$ qubits in the QC$_{bind}$ module, where the unitary operation QC$_{bind}$ is applied uniformly across all $2^m$ complexes.

$$\begin{bmatrix} O^0 \\ O^1 \\ ... \\ O^{2^m-1} \end{bmatrix}_{2^{m+n}} = \begin{bmatrix} U_{bind} & 0 & ... & 0 \\ 0 & U_{bind} & ... & 0 \\ ... & ... & ... & ... \\ 0 & 0 & ... & U_{bind} \end{bmatrix}_{2^{m+n} * 2^{m+n}} \begin{bmatrix} I^0 \\ I^1 \\ ... \\ I^{2^m-1} \end{bmatrix}_{2^{m+n}} \quad (3)$$

As shown in Figure 4, the design leverages $m + n$ qubits to simultaneously process multiple molecular interactions, showcasing the scalability and efficiency of quantum systems for SBVS.

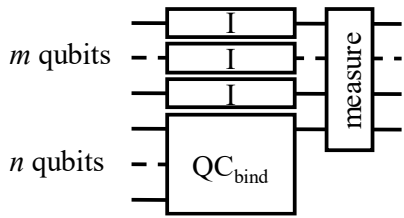

**Figure 4. Quantum Circuit for Binding Free Energy Estimation.** This conceptual quantum circuit is designed to estimate $\Delta G_{bind}$ values for multiple protein–ligand complexes in parallel. It uses $n$ qubits to encode a single complex and $m$ ancillary qubits to simultaneously evaluate up to $2^m$ binding configurations via quantum superposition. Identity gates (I) preserve the coherence of inactive qubits. The QC$_{bind}$ module performs the core binding free energy calculations, and the final measurement step projects the quantum state to obtain classical outputs**.**

**Problem Addressed.** The application of quantum computing to virtual drug screening presents numerous challenges. This study focuses explicitly on estimating protein–ligand binding free energy as an initial step toward harnessing quantum machine learning for structure-based virtual screening. In classical virtual drug screening, approaches for computing protein–ligand binding free energy generally fall into two categories: force field-based methods and machine learning models. This study adopts a machine learning approach similar to that implemented in *Gnina* [18].

Since quantum machine learning remains in its early stages, this work explores a novel design that replaces classical convolutional kernels with quantum filters (Qfilters).

However, implementing nonlinear functions, such as activation functions from classical convolutional networks, on a quantum circuit typically requires many quantum gates and qubits, significantly limiting practical feasibility. To address this, we propose and evaluate a simplified quantum model to examine the trade-off between prediction accuracy and computational efficiency.

**Quantum Circuit Architecture.** The quantum circuit architecture follows a convolutional design, wherein each convolutional layer acting on $m$ qubits is implemented as a unitary matrix of size $2^m * 2^m$. This quantum convolutional layer simultaneously processes $2^m$ data points and comprises $2^m$ channels. The resulting quantum state is a linear combination of the input states, allowing each output channel to encode features from multiple input states, although not all channels are necessarily utilized. For instance, in Figure 1a, Qfilter 1 operates on qubits 0, 3, and 4, encoding their collective quantum information into qubit 0. Similarly, Qfilter 2 acts on qubits 1, 5, and 6 and stores its output in qubits 1, while Qfilter 3 processes qubits 2, 7, and 8 with the result written to qubits 2. Finally, Qfilter 4 takes qubits 0, 1, and 2 as input and encodes their aggregated information into qubit 0, thereby integrating the information propagated through all preceding Qfilters.

Since quantum measurement yields only a single basis state per shot, multiple repetitions are required to estimate probability distributions accurately. Measuring $p$ qubits results in a distribution over $2^p$ possible outcomes. As $p$ increases, a larger number of measurement shots is required to maintain the same level of statistical precision. Reducing the number of measured qubits can significantly lower the sampling cost. Figure 1 illustrates this principle by progressively funneling quantum information through successive Qfilters into qubit 0, which is ultimately measured.

A Qfilter acting on $m$ qubits is parameterized by a unitary matrix, which generally requires $2^{2m}$ real-valued parameters. However, due to the unitarity constraint, the number of independent real-valued parameters in such a matrix is $(2^m(2^m - 1))/2$. This number increases rapidly with $m$, thereby enhancing the representational capacity of the Qfilter. At the same time, it also introduces significant computational overhead. In particular, performing SVD to obtain valid unitary matrices becomes increasingly costly. Compiling these learned unitary matrices into executable quantum gates further increases the overall circuit depth and gate count.

**Replacing Parameterized Quantum Gates with Qfilters.** In conventional quantum circuit design, parameterized quantum gates are typically employed. However, these gates rely heavily on sine and cosine functions, resulting in quantum states involving complex trigonometric products. When the loss function is defined in quantum state probabilities, it inherits these trigonometric dependencies, leading to highly non-convex loss landscapes. These trigonometric dependencies, in turn, exacerbate the barren plateau problem, characterized by vanishing gradients and stagnated training in deep quantum circuits.

Furthermore, constructing even a single Qfilter using parameterized quantum gates typically requires multiple layers of operations. Whether simulated via quantum circuit platforms or implemented through matrix multiplications in frameworks such as PyTorch, executing deep parameterized circuits is significantly more time-consuming than applying Qfilters directly as unitary matrices.

Therefore, this study adopts Qfilters as a strategic alternative during training. Qfilters enable faster iteration during parameter optimization by avoiding the need for repeated gate-level circuit execution. Once training is complete, the learned Qfilters can be decomposed into native quantum gates for implementation on hardware, preserving computational efficiency without compromising compatibility [39, 40].

**Performance of QCNN Architectures Under Noise-Free Conditions.** Under ideal, noise-free conditions, all five QCNN architectures demonstrated effective predictive performance for estimating protein–ligand $\Delta G_{bind}$. Test set RMSD values ranged from 2.23 to 2.37 kcal/mol, with corresponding PCC values between 0.653 and 0.694. Among the 9-qubit designs, the architecture shown in Fig. 1c, which uses larger 5-qubit filters, achieved the best test performance (RMSD = 2.23; PCC = 0.693), suggesting that increased filter size enhances expressiveness. However, this improvement came at the cost of a significantly larger number of parameters (2,050 total, 994 independent).

For the 12-qubit circuits, both Fig. 1f (4+4+4+3 filters) and Fig. 1g (5+5+4 filters) performed comparably. The Fig. 1f model achieved slightly higher test PCC, while the Fig. 1g model required more than twice the number of parameters (2,306 vs. 834), indicating diminishing returns in predictive accuracy relative to model complexity.

Training curves over 10,000 epochs showed consistent convergence across all learning rates. Only minor RMSD fluctuations were observed, and the variance in RMSD and PCC across different learning rates remained small. These results indicate that the QCNN architectures are robust to hyperparameter initialization and maintain stable performance under ideal simulation conditions.

**Performance Under Quantum Noise.** When quantum noise was introduced into the circuits, all QCNN architectures experienced degradation in predictive accuracy, particularly regarding RMSD. Two noise injection strategies were examined: Final Qubit Noise, where noise was applied only to the final measured qubit, and Layer-Wise Noise, where depolarizing and phase-damping noise were applied to all qubits after each QFilter operation.

Across all architectures, Final Qubit Noise resulted in moderate increases in RMSD (for example, from 2.23 to 2.38 in Figure 1c), while the PCC values remained unchanged (0.693 in Figure 1c). This indicates that although noise introduced slight perturbations in prediction magnitude, the linear correlation between predicted and true values was largely preserved.

In contrast, Layer Wise Noise led to more substantial increases in RMSD, ranging from approximately 0.3 to 0.6 kcal/mol across architectures, indicating reduced prediction accuracy. However, PCC values remained relatively stable. For instance, in Figure 1a, the test RMSD increased from 2.37 to 2.81 kcal/mol, yet the PCC changed only slightly from 0.653 to 0.652. In Figure 1b, the PCC even increased from 0.655 to 0.658, possibly due to variance smoothing or noise-induced regularization effects. These results suggest that RMSD is sensitive to quantum noise, while PCC is comparatively robust.

Despite the degradation, some architectures, such as Figure 1c and Figure 1d, retained high PCC values under both noise settings, indicating that their circuit configurations may offer greater robustness on realistic quantum hardware. The difference between the two noise models highlights the need for effective error mitigation, particularly in circuits that stack multiple QFilters.

**Implementing Qfilters with Unitary Matrices.** In this study, Qfilters were designed as unitary matrices by initially optimizing their parameters in a classical setting and subsequently applying SVD to project the resulting matrices onto valid unitary forms. This approach ensures that the Qfilters adhere to fundamental quantum mechanical principles.

Following training, the parameters were decomposed into quantum circuits, enabling efficient computation without directly constructing circuits densely packed with PQGs.

This strategy effectively mitigates the substantial simulation overhead typically associated with PQGs. Additionally, it addresses the barren plateau problem, a common issue in quantum circuit optimization, where many PQGs lead to vanishing gradients and hamper practical training. By employing unitary matrix-based Qfilters, the study achieves both computational efficiency and reliable optimization.

**Feasibility of Implementation on Real Quantum Hardware.** The first step in implementing this method on quantum hardware is to design an encoder that generates the input quantum states. This work uses Amplitude Encoding, which efficiently encodes classical data into quantum states by mapping data values to the amplitudes of a quantum superposition state. This method allows for representing high-dimensional data using a logarithmic number of qubits. However, preparing the quantum state requires complex gates, often leading to deep circuits prone to noise and decoherence.

The configuration of Qfilters, including their assignment to specific qubits and the number of Qfilters utilized, is determined by the capabilities and constraints of the quantum hardware. Trained Qfilters can be decomposed into fundamental quantum gates, such as single-qubit rotation gates (Rx, Ry, Rz) and controlled operations (CNOT, CZ). The size of each Qfilter influences the overall circuit depth, affecting computational accuracy due to susceptibility to noise and decoherence. While larger Qfilters offer greater transformation flexibility by incorporating more parameters, they also demand additional quantum gates, leading to deeper circuits and increased vulnerability to hardware-induced errors.

Given the substantial architectural variations among quantum computing platforms, adapting this method to different hardware implementations requires careful modifications. This approach provides a framework for investigating the practical feasibility of quantum computing in addressing real-world challenges.

**Conclusions**

This study evaluated the performance of various QCNN architectures for predicting protein–ligand binding free energy under both noise-free and noisy quantum simulation settings. While all architectures demonstrated reasonable predictive accuracy in ideal conditions, the introduction of quantum noise, particularly Layer-Wise Noise, led to increased RMSD values, reflecting reduced predictive precision. However, PCC remained largely stable across all noise settings, indicating that the relative ordering of predictions was preserved even when absolute errors increased. The robustness of PCC highlights the potential of QCNN models to maintain consistent qualitative performance in the presence of realistic quantum hardware noise. Architectures such as those in Figure 1c and 1d showed particularly strong resilience, suggesting that carefully designed QFilter configurations can help mitigate the effects of noise. These findings emphasize the importance of choosing noise-tolerant circuit structures and implementing effective error mitigation strategies for practical applications of quantum machine learning in structure-based virtual screening.

**Data and Software Availability**

The source code and data preprocessing scripts are publicly available at

https://github.com/peikunyang/07_QCNN_SBVS. All datasets and scripts required to reproduce this study's results are included and freely accessible.

**Reference**

[1]   Reymond J-L and Awale M 2012 Exploring chemical space for drug discovery using the chemical universe database *ACS chemical neuroscience* **3** 649-57

[2]   Polishchuk P G, Madzhidov T I and Varnek A 2013 Estimation of the size of drug-like chemical space based on GDB-17 data *Journal of computer-aided molecular design* **27** 675-9

[3]   Cheng T, Li Q, Zhou Z, Wang Y and Bryant S H 2012 Structure-based virtual screening for drug discovery: a problem-centric review *AAPS J* **14** 133-41

[4]   Willems H, De Cesco S and Svensson F 2020 Computational Chemistry on a Budget: Supporting Drug Discovery with Limited Resources: Miniperspective *Journal of medicinal chemistry* **63** 10158-69

[5]   Xiong G, Shen C, Yang Z, Jiang D, Liu S, Lu A, Chen X, Hou T and Cao D 2021 Featurization strategies for protein–ligand interactions and their applications in scoring function development *Wiley Interdisciplinary Reviews: Computational Molecular Science*  e1567

[6]   Tingle B I, Tang K G, Castanon M, Gutierrez J J, Khurelbaatar M, Dandarchuluun C, Moroz Y S and Irwin J J 2023 ZINC-22— A free multi-billion-scale database of tangible compounds for ligand discovery *Journal of chemical information and modeling* **63** 1166-76

[7]   Shirts M R, Mobley D L and Brown S P 2010 Free-energy calculations in structure-based drug design *Drug Design* **1** 61-86

[8]   Deng Y and Roux B 2009 Computations of standard binding free energies with molecular dynamics simulations *J Phys Chem B* **113** 2234-46

[9]   Gumbart J C, Roux B and Chipot C 2013 Standard binding free energies from computer simulations: What is the best strategy? *Journal of chemical theory and computation* **9** 794-802

[10]  Fu H, Chen H, Blazhynska M, Goulard Coderc de Lacam E, Szczepaniak F, Pavlova A, Shao X, Gumbart J C, Dehez F, Roux B, Cai W and Chipot C 2022 Accurate determination of protein:ligand standard binding free energies from molecular dynamics simulations *Nat Protoc* **17** 1114-41

[11]  Smith S T and Meiler J 2020 Assessing multiple score functions in Rosetta for drug discovery *PLoS One* **15** e0240450

[12]  Dittrich J, Schmidt D, Pfleger C and Gohlke H 2019 Converging a Knowledge-

Based Scoring Function: DrugScore(2018) *J Chem Inf Model* **59** 509-21

[13]   Macari G, Toti D, Pasquadibisceglie A and Polticelli F 2020 DockingApp RF: A State-of-the-Art Novel Scoring Function for Molecular Docking in a User-Friendly Interface to AutoDock Vina *Int J Mol Sci* **21** 9548

[14]   Bao J, He X and Zhang J Z H 2020 Development of a New Scoring Function for Virtual Screening: APBScore *J Chem Inf Model* **60** 6355-65

[15]   Yang C and Zhang Y 2021 Lin_F9: A Linear Empirical Scoring Function for Protein-Ligand Docking *J Chem Inf Model* **61** 4630-44

[16]   Morris G M, Huey R and Olson A J 2008 Using AutoDock for ligand-receptor docking *Curr Protoc Bioinformatics* **Chapter 8** Unit 8 14

[17]   Pagadala N S, Syed K and Tuszynski J 2017 Software for molecular docking: a review *Biophys Rev* **9** 91-102

[18]   McNutt A T, Francoeur P, Aggarwal R, Masuda T, Meli R, Ragoza M, Sunseri J and Koes D R 2021 GNINA 1.0: molecular docking with deep learning *J Cheminform* **13** 43

[19]   Trott O and Olson A J 2010 AutoDock Vina: improving the speed and accuracy of docking with a new scoring function, efficient optimization, and multithreading *J Comput Chem* **31** 455-61

[20]   Li H, Sze K H, Lu G and Ballester P J 2021 Machine-learning scoring functions for structure-based virtual screening *Wiley Interdisciplinary Reviews: Computational Molecular Science* **11** e1478

[21]   Brown B P, Mendenhall J, Geanes A R and Meiler J 2021 General Purpose Structure-Based Drug Discovery Neural Network Score Functions with Human-Interpretable Pharmacophore Maps *J Chem Inf Model* **61** 603-20

[22]   Francoeur P G, Masuda T, Sunseri J, Jia A, Iovanisci R B, Snyder I and Koes D R 2020 Three-Dimensional Convolutional Neural Networks and a Cross-Docked Data Set for Structure-Based Drug Design *J Chem Inf Model* **60** 4200-15

[23]   Kim J, Park S, Min D and Kim W 2021 Comprehensive Survey of Recent Drug Discovery Using Deep Learning *Int J Mol Sci* **22** 9983

[24]   Yang X, Wang Y, Byrne R, Schneider G and Yang S 2019 Concepts of Artificial Intelligence for Computer-Assisted Drug Discovery *Chem Rev* **119** 10520-94

[25]   Kimber T B, Chen Y and Volkamer A 2021 Deep Learning in Virtual Screening: Recent Applications and Developments *Int J Mol Sci* **22** 4435

[26]   Jiménez-Luna J, Grisoni F and Schneider G 2020 Drug discovery with explainable artificial intelligence *Nature Machine Intelligence* **2** 573-84

[27]   de Lima Marquezino F, Portugal R and Lavor C 2019 *A primer on quantum*

*computing*: Springer)

[28]  Dickens J and Salmasian H 2019 *Quantum Computing Algorithms for Applied Linear Algebra*

[29]  Loceff M 2015 *A course in quantum computing (for the community college). Foothill College*

[30]  Lipton R J and Regan K W 2014 *Quantum algorithms via linear algebra: a primer*: MIT Press)

[31]  McMahon D 2007 *Quantum computing explained*: John Wiley & Sons)

[32]  Kaye P, Laflamme R and Mosca M 2007 *An introduction to quantum computing*: Oxford University Press on Demand)

[33]  Li J, Alam M, Congzhou M S, Wang J, Dokholyan N V and Ghosh S 2021 Drug discovery approaches using quantum machine learning. In: *2021 58th ACM/IEEE Design Automation Conference (DAC)*: IEEE) pp 1356-9

[34]  Batra K, Zorn K M, Foil D H, Minerali E, Gawriljuk V O, Lane T R and Ekins S 2021 Quantum Machine Learning Algorithms for Drug Discovery Applications *J Chem Inf Model* **61** 2641-7

[35]  Li J, Topaloglu R O and Ghosh S 2021 Quantum generative models for small molecule drug discovery *IEEE Transactions on Quantum Engineering* **2** 1-8

[36]  Sunseri J, King J E, Francoeur P G and Koes D R 2019 Convolutional neural network scoring and minimization in the D3R 2017 community challenge *Journal of computer-aided molecular design* **33** 19-34

[37]  Pasquali D 2020 Simultaneous Quantum Machine Learning Training and Architecture Discovery *arXiv preprint arXiv:2009.06093*

[38]  McClean J R, Boixo S, Smelyanskiy V N, Babbush R and Neven H 2018 Barren plateaus in quantum neural network training landscapes *Nature communications* **9** 4812

[39]  Krol A M, Sarkar A, Ashraf I, Al-Ars Z and Bertels K 2022 Efficient decomposition of unitary matrices in quantum circuit compilers *Applied Sciences* **12** 759

[40]  Daskin A and Kais S 2011 Decomposition of unitary matrices for finding quantum circuits: application to molecular Hamiltonians *The Journal of chemical physics* **134**